\documentclass{bioinfo}
\copyrightyear{2022}
\pubyear{2022}

\usepackage{graphicx}
\usepackage{hyperref}
\usepackage{url}
\usepackage{tabularx}
\usepackage{amsmath}
\usepackage[ruled,vlined]{algorithm2e}

\SetCommentSty{mycommfont}
\SetKwComment{Comment}{$\triangleright$\ }{}

\usepackage{natbib}
\begin{document}
\firstpage{1}

\title[Aligning proteins to genomes with miniprot]{Protein-to-genome alignment with miniprot}
\author[Li]{Heng Li$^{1,2}$}
\address{$^1$Dana-Farber Cancer Institute, 450 Brookline Ave, Boston, MA 02215, USA,
$^2$Harvard Medical School, 10 Shattuck St, Boston, MA 02215, USA}

\maketitle

\begin{abstract}

\section{Motivation:} Protein-to-genome alignment is critical to annotating
genes in non-model organisms. While there are a few tools for this purpose, all
of them were developed over ten years ago and did not incorporate the latest
advances in alignment algorithms. They are inefficient and could not keep up
with the rapid production of new genomes and quickly growing protein databases.

\section{Results:} Here we describe miniprot, a new aligner for mapping
protein sequences to a complete genome. Miniprot integrates recent techniques
such as k-mer sketch and SIMD-based dynamic programming. It is tens of times
faster than existing tools while achieving comparable accuracy on real data.

\section{Availability and implementation:}
\href{https://github.com/lh3/miniprot}{https://github.com/lh3/miniprot}

\section{Contact:} hli@ds.dfci.harvard.edu
\end{abstract}

\section{Introduction}

Sequencing technologies have been rapidly evolving in recent years. The advent
of long-read sequencing, especially accurate long-read
sequencing~\citep{Wenger:2019ab}, have enabled high-quality genome assembly at
scale~\citep{Nurk:2020we,Cheng:2021aa,Cheng:2022aa}. After we sequence and
assemble the genome of a new species, the immediate next step is to annotate
genes.

There are three ways to annotate gene structures: \emph{ab initio} gene
prediction, aligning RNA-seq data from the same species and mapping known genes
with cross-species alignment.  While \emph{ab initio} gene prediction works
well for bacterial genomes, it is error-prone for Eukaryotic
genomes that may contain large introns. In a recent
benchmark~\citep{Scalzitti:2020wg}, all the evaluated gene finders miss
$\sim$50\% nucleotides in annotated exons and predict $\sim$50\% extra
sequences not in exons. If we have RNA-seq data, we can map short or long
RNA-seq reads~\citep{Dobin:2013kx,Li:2018ab} and reconstruct transcripts from
the alignment~\citep{Kovaka:2019wf}. This will give much more accurate gene
structures than \emph{ab initio} gene prediction. Unfortunately, RNA sequencing
adds extra cost and may miss genes lowly expressed in the tissues being
sequenced. We still rely on cross-species alignment to derive a complete gene
set and to transfer known functional annotations to the new genome.

For very closely related genomes, we can reconstruct gene structures from
whole-genome alignment~\citep{Fiddes:2018wn} or from the alignment of gene
regions~\citep{Shumate:2020ty}. These methods would not work well for genomes
at longer evolutionary distances because intron sequences are less conserved and
this will affect the quality of the alignment. Aligning the more conserved
coding regions~\citep{Li:2007aa,Gotoh:2008aa} may alleviate the issue. However,
for distantly related species, even coding nucleotide sequences are not
conserved well. Just as we almost exclusively use protein sequences to
reconstruct the phylogeny of distant homologs,
Ensembl~\citep{Aken:2016wr} and other gene annotation
pipelines~\citep{Haas:2008tv,Holt:2011tt,Bruna:2021ug} also heavily rely on
protein-to-genome alignment especially when the annotation of closely related
species is not available.

There are several protein-to-genome aligners that pinpoint exact splice sites:
GeneWise~\citep{Birney:1997vr,Birney:2004uy}, Exonerate~\citep{Slater:2005aa},
GeneSeqer~\citep{Usuka:2000vi},
GenomeThreader~\citep{DBLP:journals/infsof/GremmeBSK05},
genBlastG~\citep{She:2011aa}, ProSplign~\citep{Kapustin:2008tq} and
Spaln2~\citep{Gotoh:2008aa,Iwata:2012aa}. They all take a protein and a
nucleotide sequence as input and output spliced alignment between them.
GeMoMa~\citep{Keilwagen:2019wz} additionally requires gene structures in the
source genome as input. It aligns exons without splicing and then connects
exon alignments. As gene structures are conserved across species, this strategy
simplifies alignment and potentially reduces spurious hits but it would not
easily work for proteins from a variety of species.

Among the tools above, only Spaln2, GenomeThreader and GeMoMa are practical for
whole-genome alignment. They
can align several hundred proteins per CPU hour and may take a couple of days
to align a few hundred thousand proteins often needed to annotate a genome
without closely homology. Protein-to-genome alignment is time consuming.

It is challenging to develop a fast and accurate protein-to-genome alignment
algorithm. The core of such alignment is a dynamic programming (DP) that
jointly considers affine gap penalties, introns and frameshift. It is perhaps
the most complex DP for pairwise alignment. In addition, as we will show later,
a successful aligner functions like a gene finder and has to properly model
splice signals, which is not a trivial task, either. On top of these, we need
to fit these complex methods to an efficient implementation with modern
computing techniques. This is partly why we have over a hundred short-read
mappers~\citep{Alser:2021tk} but only three protein-to-genome mappers capable of
whole-genome alignment.

In this article, we will describe miniprot, a new protein-to-genome aligner
developed from scratch. We will demonstrate its performance and accuracy on
real data along with the few existing algorithms.

\begin{methods}
\section{Methods}

Miniprot broadly follows the seed-chain-extend strategy used by
minimap2~\citep{Li:2018ab}. It indexes the genome k-mers
in all six open reading frames (ORFs) on both
strands. During alignment, miniprot extracts k-mers on a query protein,
finds seed anchors and then performs chaining. It closes
unaligned regions between anchors and extends from terminal anchors with
dynamic programming (DP).

\subsection{Notations of strings}

For a string $T$, let $|T|$ be its length and $T[i]$, $i=1,\ldots,|T|$, be
the $i$-th symbol in $T$. $T[i,j]$, $1\le i\le j\le|T|$, is the substring
starting at $i$ and ending at $j$ inclusively. In this article, $T$ denotes the
genome sequence over the nucleotide alphabet and $P$ denotes the protein
sequence over the amino acid alphabet.

\subsection{Reduced alphabet}

There are twenty amino acids. We need at least five bits to encode each amino
acid. To encode protein sequences more compactly, we reduce the amino acid
alphabet using the SE-B(14) scheme by ~\citet{Edgar:2004aa}, except that we
merge N and D. More exactly, we map amino acid groups to integers as
follows: A$\to$0, ST$\to$1, RK$\to$2, H$\to$3, ND$\to$4, EQ$\to$5, C$\to$6,
P$\to$7, G$\to$8, IV$\to$10, LM$\to$11, FY$\to$12, W$\to$13, $\ast$$\to$14 and
X$\to$15, where $\ast$ denotes the stop codon and X denotes an amino acid.

Under this encoding, if two amino acid groups only differ at the lowest bit
(e.g. group `A' and `ST'), the two groups tend to be similar. We may flip the
lowest bit of an integer to generate more seeds and thus to increase the
seeding sensitivity. We did not use this strategy as miniprot seems reasonably
sensitive on real data.

\subsection{Indexing the genome}

Miniprot only indexes a subset of k-mers in the genome. Suppose $\phi(a)$ maps
an amino acid $a$ to a 4-bit integer with the scheme described above. The
integer encoding of a $k$-long protein sequence $P$ can be recursively defined as
$\phi(P)=\phi(P[1,k-1])\times16+\phi(P[k])$. $\phi(P)$ has $4k$ bits. Let
$B=\psi(\phi(P))$ where $\psi(\cdot)$ is an invertible integer hash
function~\citep{Li:2016aa} over $[0,2^{4k})$. Then $B$ is also an integer with
$4k$ bits. By default, miniprot only indexes $B$ if the lowest bit of $B$ is 0.
We thus sample 50\% of k-mers in average with a high-quality hash function
$\psi(\cdot)$.

Internally, miniprot treats each genome sequence and its reverse complement as
two independent sequences. It enumerates all ORFs of 30 amino acids or longer
and samples 6-mers from translated ORFs with the strategy above. For
each selected k-mer $R$ at position $x$, miniprot stores
$(\psi(\phi(R)), \lfloor x/256\rfloor)$ in a hash table with the key being
$\psi(\phi(R))$ and the value being an array of positions. We do not retain the
base resolution at the indexing step such that we can use 32-bit integers to
store positions for a genome up to $2^{39}$ ($=2^{32}\cdot 256/2$) base pairs
in size. Without binning, miniprot would have to use 64-bit integers to store
positions in a human genome, which would double the index size.

\subsection{Chaining}

The miniprot chaining algorithm is similar to the minimap2 algorithm.
However, because the miniprot index does not keep the exact genome positions,
the gap size calculation needs to be modified. For completeness, we will
describe the full chaining equation here.

Let 2-tuple $(x,y)$ denote a seed match, also known as an anchor, between binned position
$x$ on the genome and residue position $y$ on the protein. Suppose $(x_i,y_i)$
and $(x_j,y_j)$ are two anchors with $x_i\le x_j$ and $y_i<y_j$. The
minimum possible gap size between the two anchors, in the unit of base pair,
can be calculated by
\begin{equation}\label{eq:gap}
g(i,j)=\left\{\begin{array}{ll}
3\Delta y-256(\Delta x-1) & \mbox{if $3\Delta y<256(\Delta x-1)$}\\
3\Delta y-256(\Delta x+1) & \mbox{if $3\Delta y>256(\Delta x+1)$}\\
0 & \mbox{otherwise}
\end{array}\right.
\end{equation}
with $\Delta x=x_j-x_i$ and $\Delta y=y_j-y_i$. When $g(i,j)=0$, we do not know
if there is a gap due to binning. Meanwhile, $g(i,j)>0$ indicates a definitive
insertion to the genome and $g(i,j)<0$ indicates a definitive deletion.

Given a list of anchors sorted by genomic position $x$, let $f(j)$ be the
maximal chaining score up to the $j$-th anchor in the list. $f(j)$ can be
calculated with
\begin{equation}
f(j)=\max\big\{\max_{1\le i<j}\{f(i)+\alpha(i,j)-\gamma(g(i,j))\},k\big\}
\end{equation}
where $k$ is the k-mer length (6 amino acids by default), $g(i,j)$
is calculated by Eq.~(\ref{eq:gap}) and $\alpha(i,j)=\min\{y_j-y_i,k\}$ is the
number of matching residues between the anchors. The gap penalty function
$\gamma(\cdot)$ is
\begin{equation}\label{eq:gap-pen}
\gamma(g)=\left\{\begin{array}{ll}
0 & \mbox{if $g=0$}\\
\infty & \mbox{if $|g|\ge G$}\\
g/3+\beta\log_2(g+1) & \mbox{if $0<g<G$}\\
\min\{|g|/3,\beta\log_2(|g|+1)\} & \mbox{if $-G<g<0$}\\
\end{array}\right.
\end{equation}
Here $G$ is the maximum intron size (200 kb by default) and $\beta$ is the
weight of the logarithm gap penalty (0.75 by default).  The logarithm term
allows miniprot to join exons over introns.

After the initial round of chaining for each protein, miniprot selects the top
30 chains and performs another round of chaining in local regions around these
top chains. In the second round, miniprot indexes all 5-mers on both the
protein and the genome subsequences without binning. This finds better chains
and retains the base resolution of each anchor. Miniprot uses $g'(i,j)=3\Delta
y-\Delta x$ to compute gap lengths and applies the same gap penalty
Eq.~(\ref{eq:gap-pen}) during chaining.

\subsection{Residue alignment with dynamic programming}

Miniprot uses DP to close gaps between anchors in chains and to extend from
terminal anchors. The DP aims to find gaps, frameshift and splicing at the
same time as is demonstrated as follows (``Geno'' for the genome sequence,
``Tran'' for the translated protein sequence in the alignment and ``Prot'' for
the query protein sequence):
\begin{verbatim}
Geno: GAGGCC---CGCTCACCgt...agCACAAGCGCTATAGCCTAC
Tran: E..A..---R..S..P.       .T..$$R..Y..+A..Y..
      |  |     |     |         |    |  |   |  |
Prot: E  A  F  R  -  P         T  E R  Y   A  Y
\end{verbatim}
In this example, symbol ``{\tt \$}'' denotes frameshift substitutions and
``{\tt +}'' denotes frameshift insertions. We will explain their differences
later. In this section, we will first review the AE86 DP formulation for affine
gap cost~\citep{Altschul:1986aa}, and then derive the DP equation for
protein-to-genome alignment.

\subsubsection{DP with affine gap cost}

Under the affine gap cost, a gap of length $g$ costs $q+e\cdot g$. A direct
formulation of the DP looks like
\begin{equation}\label{eq:durbin}
\left\{\begin{array}{lll}
M_{ij}&=&\max\{ M_{i-1,j-1}, I_{i-1,j-1}, D_{i-1,j-1} \} + s(i,j) \\
I_{ij}&=&\max\{ M_{i,j-1} - q, I_{i,j-1}, D_{i,j-1} - q \} - e\\
D_{ij}&=&\max\{ M_{i-1,j} - q, D_{i,j-1} - q, D_{i-1,j} \} - e\\
\end{array}\right.
\end{equation}
where `$M$' represents the matching state, `$I$' the insertion state, `$D$` the
deletion state and $s(i,j)$ gives the score between the residue at position $i$
on the target sequence and the residue at position $j$ on the query. If we
define
$$
H_{ij} = \max\{M_{ij},I_{ij},D_{ij}\}
$$
Eq.~(\ref{eq:durbin}) becomes
\begin{equation}\label{eq:ae86}
\left\{\begin{array}{lll}
I_{ij}&=&\max\{ H_{i,j-1} - q, I_{i,j-1} \} - e\\
D_{ij}&=&\max\{ H_{i-1,j} - q, D_{i-1,j} \} - e\\
H_{ij}&=&\max\{ H_{i-1,j-1} + s(i,j), I_{ij}, D_{ij} \} \\
\end{array}\right.
\end{equation}
Eq.~(\ref{eq:ae86}) is the AE86 formulation. It invokes fewer comparisons. When
there are more states, AE86 may save more comparisons and simplify the DP
equation.

\subsubsection{DP for protein-to-DNA alignment}

In a similar manner, we can derive the DP for protein-to-DNA alignment,
allowing frameshifts but not splicing:
\begin{equation}\label{eq:fs}
\left\{\begin{array}{lll}
I_{ij}&=&\max\{ H_{i,j-1} - q, I_{i,j-1} \} - e \\
D_{ij}&=&\max\{ H_{i-3,j} - q, D_{i-3,j} \} - e \\
H_{ij}&=&\max\{ H_{i-3,j}+s(i,j), I_{ij}, D_{ij}, H_{i-1,j-1}-f, \\
       && H_{i-2,j-1}-f, H_{i-1,j}-f, H_{i-2,j}-f \} \\
\end{array}\right.
\end{equation}
It is similar to Eq.~(\ref{eq:ae86}) except for codon phase transitions with a
penalty of $f$. We have two types of frameshift. The first type is created by
inserting one or two bases to the DNA sequence (symbol `{\tt +}' in the example
above) and the second type by deleting one or two bases in a codon (`{\tt \$}'
in the example). These are modeled by the four $H_{\cdot,\cdot}$ terms on the
last line of Eq.~(\ref{eq:fs}).  This equation is broadly similar
to~\citet{Zhang:1997tq}.

\subsubsection{DP for protein-to-genome alignment}

When aligning proteins to genomes, we need to keep phases through introns. We
add three additional states, $A$, $B$ and $C$, for phase-0, phase-1 and phase-2
introns, respectively.  Our final formulation is
\begin{equation}\label{eq:full}
\left\{\begin{array}{lll}
I_{ij}&=&\max\{ H_{i,j-1} - q, I_{i,j-1} \} - e \\
D_{ij}&=&\max\{ H_{i-3,j} - q, D_{i-3,j} \} - e \\
A_{ij}&=&\max\{ H_{i-1,j} - r - d(i-1), A_{i-1,j} \} \\
B_{ij}&=&\max\{ H_{i-1,j-1} - r - d(i), B_{i-1,j} \} \\
C_{ij}&=&\max\{ H_{i-1,j-1} - r - d(i+1), C_{i-1,j} \} \\
H_{ij}&=&\max\{ H_{i-3,j}+s(i,j), I_{ij}, D_{ij}, H_{i-1,j-1}-f, \\
       && H_{i-2,j-1}-f, H_{i-1,j}-f, H_{i-2,j}-f, \\
	   && A_{ij}-a(i), B_{ij}-a(i-2), C_{ij}-a(i-1) \}
\end{array}\right.
\end{equation}
where $r$ is cost of an intron, and $d(\cdot)$ and $a(\cdot)$ model splice
signals. The great majority of introns start with ${\tt GT}$ and end with ${\tt
AG}$ across all species. For a simple model, we may define:
$$
d(i)=\left\{\begin{array}{ll}
0 & \mbox{if $T[i+1,i+2]={\tt GT}$}\\
p & \mbox{otherwise}\\
\end{array}\right.
$$
and
$$
a(i)=\left\{\begin{array}{ll}
0 & \mbox{if $T[i-1,i]={\tt AG}$}\\
p & \mbox{otherwise}\\
\end{array}\right.
$$
This still allows non-${\tt GT}$-${\tt AG}$ splicing but penalizes such introns
by cost $p$. We will describe a more sophisticated model in the next section.

It is worth noting that when the DP transitions from state $H$ to $B$ at
position $i$, the phase-1 intron $B$ starts at $i+1$; when the DP
transitions from $B$ to $H$ at $j$, the intron ends at $j-2$. The DP ignores
the split codon bridging the two exons around the phase-1 intron. Phase-2
intron state $C$ is treated similarly. Not scoring split codons is a weakness
of our equation.

Though not explicitly derived from a Hidden Markov Model (HMM),
Eq.~(\ref{eq:full}) is similar to the Viterbi decoding of the 6-state HMM
employed by GeneWise~\citep{Birney:2004uy} and Exonerate~\citep{Slater:2005aa}.
To that end,  our formulation should have comparable accuracy to the two older
aligners if they are parameterized the same way.

We implemented Eq.(\ref{eq:full}) with striped DP~\citep{Farrar:2007hs}.
We used 16-bit integers to keep scores and achieved 8-way parallelization
for x86\_64 CPUs with SSE2 or ARM64 CPUs with the NEON instruction set.
Our implementation is over 50 times times faster than GeneWise and Exonerate in
their exact mode.

\subsection{Splice models}

We observed that under distant homology, the splice model may have a large
influence on the junction accuracy, confirming~\citet{Iwata:2012aa}.

The most common splice pattern in all species is ${\tt GT}$-${\tt AG}$ with
${\tt GT}$ at the donor site (5'-end of an intron) and ${\tt AG}$ at the
acceptor site (3'-end of an intron). We occasionally see ${\tt GC}$-${\tt AG}$
and ${\tt AT}$-${\tt AC}$ at $\sim$1\% frequency in total~\citep{Sheth:2006vg}.
Among the ${\tt GT}$-${\tt AG}$ class, we more often observe ${\tt GTR}$-${\tt
YAG}$ from yeasts to mammals~\citep{Irimia:2008aa}, where ${\tt R}$ denotes
${\tt A}$ or ${\tt G}$ and ${\tt Y}$ denotes ${\tt C}$ or ${\tt T}$.

The default miniprot splice model considers the signals above. Using human data
from~\citet{Sibley:2016vh}, we estimated that 99.81\% of acceptor sites are ${\tt
AG}$ and only 0.10\% are ${\tt AC}$. In the BLOSUM
scaling~\citep{Henikoff:1992tk}, an ${\tt AC}$ acceptor site would be penalized
by $2\log_2 99.81/0.10\approx 20$. We can adapt this approach for three bases at
either the donor or the acceptor sites. In our final model,
$$
d(i)=\left\{\begin{array}{ll}
0 & \mbox{if $T[i+1,i+3]={\tt GTA}$ or ${\tt GTG}$}\\
8 & \mbox{if $T[i+1,i+3]={\tt GTC}$ or ${\tt GTT}$}\\
15 & \mbox{if $T[i+1,i+2]={\tt GC}$}\\
21 & \mbox{if $T[i+1,i+2]={\tt AT}$}\\
30 & \mbox{otherwise}\\
\end{array}\right.
$$
and
$$
a(i)=\left\{\begin{array}{ll}
0 & \mbox{if $T[i-2,i]={\tt CAG}$ or ${\tt TAG}$}\\
8 & \mbox{if $T[i-2,i]={\tt AAG}$ or ${\tt GAG}$}\\
21 & \mbox{if $T[i-1,i]={\tt AC}$}\\
30 & \mbox{otherwise}\\
\end{array}\right.
$$

In mammals and even \emph{Drosophila}, the last exon base adjacent to a donor site
is more often a ${\tt G}$ and we often see a poly-pyrimidine (i.e. ${\tt C}$ or
${\tt T}$) sequence close to an acceptor site. Our human splice model
considers these signals. It is also applicable to species with the sequence
features above, including \emph{Drosophila}.

Exonerate uses a position-specific weight matrix over $\sim$10 positions to
model splice sites. Spaln2 additionally considers branching sites and provides
pre-trained models for a variety of species. Miniprot adopts a relatively
simple model with fewer parameters. This makes the model more general but may
affect the accuracy of alignment. We are considering a second pass with a
splice model trained from the first pass. This strategy is often used in
mainstream gene finders~\citep{Bruna:2021ug}.

\subsection{Avoiding pseudogenes}

If a spliced gene has an unspliced pseudogene, the unspliced pseudogene may get
a better DP score because the alignment to the pseudogene does not pay intron
penalties. To reduce the effect of pseudogenes, miniprot recalculates a DP
score between the query protein and the translated coding region without
introns. In addition, miniprot further penalizes single-exon alignment by
intron open score $r$ in Eq.(\ref{eq:full}) in case a pseudogene is aligned
better by chance.

\end{methods}

\section{Results}

\subsection{Evaluation datasets}

To evaluate the accuracy of miniprot, we collected the protein-coding gene
annotations of various species: human (\emph{Homo sapiens}) from Gencode v41,
mouse (\emph{Mus musculus}) from Gencode M30, zebrafish (\emph{Danio rerio})
and fruit fly (\emph{Drosophila melanogaster}) from Ensembl v107 and mosquito
(\emph{Anopheles gambiae}) from Ensembl metazoan v54. We selected the longest
protein for each gene to reduce redundant sequences. We mapped zebrafish and
mouse proteins to the primary assembly of the human reference genome GRCh38 and
mapped mosquito proteins to the Drosophila BDGP6 genome.

\subsection{Evaluated tools}

To evaluate what aligners can map proteins to a whole genome, we randomly
sampled 1\% of zebrafish proteins and mapped with various aligners. Only
miniprot-0.7, Spaln2-2.4.13c~\citep{Iwata:2012aa}, GeMoMa-1.9~\citep{Keilwagen:2019wz}
GenomeThreader-1.7.3~\citep{DBLP:journals/infsof/GremmeBSK05} could finish the
alignment in an hour.  GenomeThreader found less than 30\% of coding regions in
Spaln2 or miniprot alignment. It is not sensitive enough for the human-fish
divergence and thus not evaluated on the full dataset. We also evaluated
MetaEuk-r6~\citep{Levy-Karin:2020to}. Although MetaEuk does not find exact
splice sites, it may be still useful for locating coding
regions~\citep{Manni:2021ww}.

When running Spaln2, we applied option ``-Q7 -T\# -yS -LS -yB -yZ -yX2'' where
``\#'' specifies the species-specific splice model. Option ``-LS'' enables
local alignment and yields sligtly better alignment overall. Option ``-yB -yZ
-yX2'' apparently has no effect for human-zebrafish alignment but it greatly
improves the junction accuracy of the fly-mosquito alignment. We let Spaln2
choose the maximum intron and gene size automatically. Miniprot finds introns
up to 200 kbp in length by default. We changed this value to 50 kbp for
fly-mosquito alignment.

We ran GeMoMa with MMseqs2~\citep{Steinegger:2017aa} as the underlying
engine. We evaluated the best unfiltered alignment of each protein as GeMoMa
discarded most alignments in the final output. We tried to specify the maximum
intron length to 200 kb but GeMoMa took more than 320 GB memory and was killed
on our cluster. We thus used 50 kb for all alignment. GeMoMa crashed for the
human-mouse dataset at the splice alignment step.

In principle, we could localize a protein with a whole-genome mapper above and
then run GeneWise, GeneSeqer and Exonerate in local regions. However, this
would not evaluate mapping accuracy. In addition, \citet{Iwata:2012aa} have
already shown Spaln2 outperformed these older tools. We thus ignored them in
evaluation.

\subsection{Evaluating protein-to-genome alignment}

\begin{table*}[!tb]
\processtable{Evaluating protein-to-genome alignment}
{\label{tab:eval}
\begin{tabular*}{\textwidth}{@{\extracolsep{\fill}}lrrrrrrrrrrrr}
\toprule
Genome species     & human   & human   & human   & human   & human   & human   & human   & human   &fruit fly&fruit fly&fruit fly \\
Protein species    &zebrafish&zebrafish&zebrafish&zebrafish&zebrafish&zebrafish& mouse   & mouse   & mosquito& mosquito& mosquito \\
Aligner            & miniprot& miniprot& Spaln2  & Spaln2  &  GeMoMa & MetaEuk & miniprot& Spaln2  & miniprot& Spaln2  &   GeMoMa \\
Splice model       & human   & general & human   & default &     N/A &     N/A & human   & human   & human   &fruit fly&      N/A \\
\midrule
Elapsed time (sec) &     267 &     257 &  10,708 &  11,097 &   8,718 &   2,518 &     164 &   3,736 &      34 &   2,528 &    3,378 \\
Peak RAM (GB)      &    21.8 &    22.5 &     9.3 &     8.9 &   146.9 &    22.0 &    15.3 &     5.6 &     3.9 &     2.7 &     53.5 \\
\# proteins        &  25,007 &  25,007 &  25,007 &  25,007 &  25,007 &  25,007 &  21,844 &  21,844 &  13,094 &  13,094 &   13,094 \\
\# multi-exon      &  16,866 &  17,104 &  13,643 &  13,854 &  23,109 &  10,220 &  17,065 &  16,865 &   6,675 &   5,630 &   11,420 \\
\# predicted junc. & 157,918 & 161,295 & 151,388 & 209,312 & 204,764 &  78,639 & 167,446 & 171,241 &  22,614 &  27,582 &   43,203 \\
\# non-ovlp. junc. &     482 &     802 &   1,206 &  15,658 &   5,712 &     183 &     330 &     852 &     488 &     877 &    5,997 \\
\# confirmed junc. & 145,545 & 144,734 & 136,916 & 129,645 & 153,781 &   5,710 & 162,675 & 162,551 &  19,774 &  22,606 &   25,513 \\
\% confirmed junc. & 92.16\% & 89.73\% & 90.44\% & 61.94\% & 75.10\% &  7.26\% & 97.15\% & 94.93\% & 87.44\% & 81.96\% &  59.05\% \\
\% base SN         & 63.11\% & 63.16\% & 57.16\% & 55.74\% & 67.02\% & 47.71\% & 90.10\% & 88.62\% & 56.10\% & 50.21\% &  65.08\% \\
\% base SP         & 95.43\% & 94.91\% & 95.11\% & 86.75\% & 88.70\% & 91.84\% & 97.26\% & 95.27\% & 96.69\% & 97.35\% &  96.10\% \\
\botrule
\end{tabular*}
}{Protein-to-genome alignments are compared to the annotated genes in ``Genome
species''. ``\# multi-exon'' gives the number of proteins mapped with multiple exons.
A splice junction (junc.) is confirmed if it is annotated in ``Genome species''
with exact boundaries; is non-overlapping (non-ovlp.) if the intron in the
junction is not overlapping with annotated introns. ``\% confirmed junc.'' is
the percentage of predicted junctions that are confirmed. Base
sensitivity (base SN) is the fraction of annotated coding regions on the
longest transcripts that are covered by alignments. Base specificity (base SP)
is the fraction of genomic bases in alignments that are covered by annotated
coding regions.}
\end{table*}

We aligned zebrafish proteins to GRCh38 using all tools (Table~\ref{tab:eval}).
With human-specific splice models, miniprot is slightly more accurate than
Spaln2 on most metrics. Nonetheless, for proteins mapped by both miniprot and
Spaln2, Spaln2 could find more correct junctions. Looking at proteins Spaln2
aligned better, we observed Spaln2 is more sensitive to small introns and small
exons, while
miniprot tends to merge them to adjacent alignments. We speculate this may be
caused by two factors. First, Spaln2 uses a more sophisticated splice model and
may be putting more weight on splice signals than residue alignment. It may
create an intron even if the alignment is weak. Second, the Spaln2 developers
observed that heuristics may be doing better than strict DP around short
introns or exons. In one case, Spaln2 correctly created an exon with one amino
acid. Miniprot under the current setting would not produce such an alignment.
On the other hand, while Spaln2 found more correct junctions for proteins
mapped by both miniprot and Spaln2, it also produced more false junctions
related to small exons and introns. It is not clear to us what is the best
balance point.

GeMoMa is more sensitive than both miniprot and Spaln2, finding more junctions
and more annotated coding regions. It however has lower junction accuracy. We
could tune miniprot for increased sensitivity but we decided to keep the
current behavior as the additional alignments are less accurate. MetaEuk was
not designed to find exact splice junctions, as is expected.

For the human-mouse alignment, miniprot is again slightly better than Spaln2.
GeMoMa crashed. On the more challenging fly-mosquito dataset, Spaln2 has
higher junction sensitivity and higher base specificity than miniprot. GeMoMa
continues to have the highest sensitivity but lower junction accuracy and base
specificitiy.

Miniprot is over 30 times faster than Spaln2 and GeMoMa. The performance gap between miniprot and Spaln2
increases with divergence. This is potentially because Spaln2 has to invoke DP
through introns more often when it does not see overlapping high-scoring
segment pairs (HSPs) and cannot initiate ``sandwich DP''~\citep{Wu:2005vn} to
skip introns. With a much faster DP implementation, miniprot can afford to
align through all introns regardless of sequence divergence. It thus has more
stable performance. Always aligning through introns might be a contributing
factor to the higher specificity of miniprot even though Spaln2 has a more
careful algorithm.

\section{Discussions}

Miniprot is a fast protein-to-genome aligner comparable to existing tools in
accuracy. Its primary use case is to assist gene annotation. At present, the
Ensembl pipeline~\citep{Aken:2016wr} still relies on
GeneWise~\citep{Birney:2004uy} and Exonerate~\citep{Slater:2005aa}.
MAKER2~\citep{Cantarel:2008ue,Holt:2011tt} calls Exonerate.
BRAKER2~\citep{Bruna:2021ug} integrates Spaln2~\citep{Iwata:2012aa} and
depends on ProtHint~\citep{Bruna:2020vy} which also optionally invokes Spaln2.
As older protein-to-genome aligners are relatively inefficient, researchers
often use faster approximate methods to localize proteins and then apply these
aligners. Now with miniprot, we can perform approximate mapping and exact
splice alignment in one go and thus simplify existing pipelines. In addition,
when there are closely related species, miniprot could find 90\% coding regions
in minutes (see ``base SN'' on the human-mouse dataset in Table~\ref{tab:eval}).
It could also be useful for evaluating de novo assemblies~\citep{Manni:2021ww}.

Miniprot would not replace full-pledge gene
annotation pipelines such as BRAKER2~\citep{Bruna:2021ug}.  Miniprot aligns
each protein independently. When multiple proteins are mapped to the same
locus, miniprot is unable to merge identical gene models or resolve conflicts
between alignments. In addition, although miniprot has a realistic splice
model, it is not as sophisticated as the BRAKER2 model and is not trained on
the target genome. More importantly, BRAKER2 has an \emph{ab initio} gene
prediction component and may find genes with weak homology to the input
proteins. We are considering to improve our splice model and to develop a
separate tool to reconcile overlapping gene models in simple cases. This may
provide a convenient annotation pipeline when closely related species are
available.

We are evaluating the possibility to support HMMER
profiles~\citep{Eddy:2011tg} as queries. As a HMMER profile summarizes a gene
family from multiple species, it may reduce the number of queries and improve
the sensitivity of miniprot for distant homologs. There are two algorithmic
challenges: seeding and alignment.  For seeding, we could generate seeds from
the most probable protein or sample multiple seeds directly from the profile;
for alignment, we could introduce position-specific substitution cost and gap
cost. Nonetheless, the exact solution to these challenges and how much HMMER
profiles may improve the alignment remain unknown.

The Vertebrate Genome Project~\citep{Rhie:2021ug}, the Darwin Tree of Life
project, the Earth Biogenome Project~\citep{Lewin:2018ve} and many other sequencing efforts are
going to sequence hundreds of thousands of species to the reference quality in
coming years. The annotation of these genomes is as important as the assembly.
While we have seen rapid evolution of sequencing technologies and assembly
algorithms in recent years, we still heavily rely on core annotation tools
developed more than a decade ago. Miniprot is one effort to replace the
protein-to-genome alignment step with modern techniques. We look forward to
renewed development of other core annotation tools from the community.

\section*{Acknowledgements}

We thank Fergal Martin, Richard Durbin and Ewan Birney for helpful discussions
on the miniprot algorithm.

{\bf Funding:} NHGRI R01HG010040 and Chan-Zuckerberg Initiative

\bibliography{miniprot}

\end{document}